# 3DInvNet: A Deep Learning-Based 3D Ground-Penetrating Radar Data Inversion

Qiqi Dai, Yee Hui Lee, *Senior Member, IEEE*, Hai-Han Sun, Genevieve Ow,
Mohamed Lokman Mohd Yusof, and Abdulkadir C. Yucel, *Senior Member, IEEE*

*Abstract*—The reconstruction of the 3D permittivity map from ground-penetrating radar (GPR) data is of great importance for mapping subsurface environments and inspecting underground structural integrity. Traditional iterative 3D reconstruction algorithms suffer from strong non-linearity, ill-posedness, and high computational cost. To tackle these issues, a 3D deep learning scheme, called 3DInvNet, is proposed to reconstruct 3D permittivity maps from GPR C-scans. The proposed scheme leverages a prior 3D convolutional neural network with a feature attention mechanism to suppress the noise in the C-scans due to subsurface heterogeneous soil environments. Then a 3D U-shaped encoder-decoder network with multi-scale feature aggregation modules is designed to establish the optimal inverse mapping from the denoised C-scans to 3D permittivity maps. Furthermore, a three-step separate learning strategy is employed to pre-train and fine-tune the networks. The proposed scheme is applied to numerical simulation as well as real measurement data. The quantitative and qualitative results show the networks' capability, generalizability, and robustness in denoising GPR C-scans and reconstructing 3D permittivity maps of subsurface objects.

## I. Introduction

GROUND-PENETRATING radar (GPR) has been widely used in geophysical exploration and civil engineering applications due to its low-cost and non-destructive characteristics. The 3D permittivity maps reconstructed from the data obtained by GPR are highly useful to get information about the subsurface objects, such as their shapes, sizes, positions, orientations, and permittivities. Such information would be highly beneficial for subsurface object identification, geophysical defect detection, and health monitoring of the subsurface utilities and living entities (such as tree roots) [1].

There are several traditional algorithms to reconstruct 3D subsurface images from GPR C-scans, such as the back-projection algorithm (BPA) [2], Kirchhoff migration [3], and reverse time migration (RTM) [4]. However, these migration algorithms can only provide an approximation to the objects' positions and shapes but cannot reconstruct their permittivity maps, while the permittivity information is of importance for object identification and health examination. To reconstruct the subsurface permittivity map, a full-wave inversion (FWI) algorithm is proposed in [1], [5], [6]. This algorithm reconstructs the permittivity maps of the subsurface structures from GPR data via a non-linear least-squared optimization process. However, due to its high computational cost for processing 3D GPR data, most of the current FWI studies only focus on 2D cases. To date, there are only two studies that apply 3D FWI to reconstruct subsurface permittivity or conductivity distribution from GPR data. In [7], an FWI algorithm with total variation (TV) regularization and Hessian approximation was adopted. It takes about a week to generate one 3D permittivity map of a small 3D domain. To improve the efficiency of 3D FWI, a modified total variation (MTV) regularization scheme was proposed in [8]. However, the inversion process still took about 16 hours to produce a 3D permittivity map. Besides, only subsurface objects with simple and regular geometries were considered while applying FWI algorithms. The high computational complexity and restricted generalizability of the FWI algorithms greatly limit their applications in reconstructing complex 3D subsurface scenarios.

In recent years, deep learning techniques have been introduced to solve inverse scattering and electromagnetic (EM) imaging problems [9]-[11]. These studies, including the fully data-driven and physics-assisted learning approaches, have demonstrated the success of deep learning in the field of EM imaging. For example, in [12], to reconstruct relative permittivity profiles from scattered fields, U-Net architecture [13] was introduced with three different schemes, including direct inversion, backpropagation, and dominant current schemes. In [14], the connection between a deep neural network (DNN) architecture and iterative methods for nonlinear EM inverse scattering was formed and then three-cascaded convolutional neural networks (CNNs) were used to process the input complex-valued images to determine the permittivity map of the investigation domain. In [15], a complex-valued deep

This work was supported by the Ministry of National Development Research Fund, National Parks Board, Singapore (*Corresponding authors: Yee Hui Lee; Abdulkadir C. Yucel*).

Q. Dai, Y. H. Lee, and A. C. Yucel are with the School of Electrical and Electronic Engineering, Nanyang Technological University, Singapore 639798 (e-mails: daiq0004@e.ntu.edu.sg, {eyhlee, acyucel}@ntu.edu.sg).

H. -H. Sun is with the School of Mechanical and Aerospace Engineering, Nanyang Technological University, Singapore, and School of Engineering, University of Tasmania, Australia (e-mail: haihan.sun199403@gmail.com).

G. Ow and M. L. M. Yusof are with the National Parks Board, Singapore 259569 (e-mails: {genevieve_ow, mohamed_lokman_mohd_yusof }@nparks.gov.sg).

convolutional encoder-decoder network was employed to extract feature fragments from received scattered field data and retrieve the dielectric contrasts of the scatterers. Combined with a distorted-Born scheme for reconstructing a rough image of the unknown object in an inhomogeneous background, a generative adversarial network (GAN) was carefully designed in [16] to output the refined permittivity map. These works have proved the high efficiency and accuracy of deep learning techniques for solving general EM inverse problems.

Deep learning-based methods have been further been investigated for GPR applications [17]-[18], especially for GPR inverse problems such as GPR image classification, signature recognition, subsurface object detection, and restoration of the objects' properties [19]-[23]. These techniques have demonstrated their capability to learn the non-linear mapping between input GPR data and the desired outputs. DNNs have been investigated to reconstruct 2D subsurface permittivity maps from GPR B-scans [24]-[27]. In [24], three types of DNNs, including the encoder-decoder [28], U-Net [13], and conditional GAN [29], were introduced to reconstruct the 2D subsurface permittivity maps from B-scans. In [25], the U-Net architecture with instance normalization was employed to generate 2D permittivity distribution from GPR data. To tackle the challenges of reconstructing the complex permittivity maps of tunnel linings from B-scans, a trace-to-trace encoder-decoder network, called GPRInvNet, was proposed in [26]. Based on that, an improved permittivity inversion network PINet was proposed in [27] to address attenuation-induced issues. However, all the works mentioned above only focus on the 2D-domain permittivity map restoration, which can only contribute to the sectional information of the subsurface scenarios. Information such as the orientation and shape of the object cannot be retrieved especially when the subsurface object is complex, and one cross-section is not sufficient to generate the full profile of the object. Furthermore, the full-wave simulations used for 2D inversion cannot capture the actual EM phenomena in 3D real world. In 2D modeling, while the scattering is assumed to be invariant in one coordinate direction, the line sources used as excitations are assumed to be infinitely long and the polarization can only be linear. However, these are not the cases in realistic 3D modeling and the results obtained by 2D modeling could be totally different from the actual response of 3D modeling [7]. Reconstructing 3D permittivity maps via 3D full-wave simulations can address the limitations of 2D reconstruction. However, it cannot be directly achieved via the aforementioned works. Therefore, it is imperative to investigate deep learning methods for reconstructing 3D subsurface permittivity maps.

One major challenge for subsurface permittivity map reconstruction is the interference of noise patterns such as the direct coupling, reflections from the ground, and environmental noise with object reflections in the GPR data. In the past years, many traditional methods have been proposed for GPR data denoising, including mean subtraction [30]-[32], singular value decomposition (SVD) [33]-[34], principal component analysis (PCA) [35]-[36], and non-negative matrix factorization (NMF) [37]-[38]. However, the denoising performance of these signal processing algorithms degrades significantly in complicated heterogeneous environments, especially when the field strengths or shapes of noise patterns are similar to those of objects' reflection signatures. To enhance the accuracy and efficiency of GPR data denoising, deep learning-based schemes have been proposed [23], [39]-[43]. In [23], [39], and [40], DNNs were used to identify regions of interest (RoIs) that contain hyperbolae or to generate a hyperbolic binary mask to highlight the object reflection area. In [41]-[43], 2D encoder-decoder networks and U-Nets were employed to predict clutter-free B-scan images that only contain object reflections. Their experimental results demonstrate the effectiveness of deep learning techniques in eliminating environmental noise and preserving object reflections. The comparative studies in [42]-[43] prove the superiority of deep learning-based methods over the traditional algorithms on GPR data denoising.

In this work, we propose a deep learning scheme, called 3DInvNet, that reconstructs subsurface 3D permittivity maps from GPR C-scans with prior denoising. To the best of our knowledge, this is the first work that performs 3D subsurface scenario reconstruction from GPR data via a deep learning scheme. The main contributions of our scheme compared to the existing deep learning-based GPR detection/2D reconstruction schemes are summarized as follows.

1) A prior 3D denoising network, called Denoiser, is designed to suppress noise in GPR C-scans due to heterogeneous soil environments. Different from complicated networks in [41]-[43], the Denoiser adopts a small 3D CNN with residual learning and feature attention mechanism to extract subsurface objects' reflection signatures in noisy C-scans effectively.
2) After denoising, a 3D U-shaped encoder-decoder network, called Inverter, is designed to map the denoised C-scans predicted by the Denoiser into subsurface 3D permittivity maps. Instead of using conventional convolutions with single scale in [44], multi-scale feature aggregation modules are proposed to extract the features of multiple objects with various properties.
3) A three-step separate learning strategy is used to pre-train and fine-tune the Denoiser and Inverter.

Test results clearly show that the proposed scheme is capable of denoising C-scans and then reconstructing 3D permittivity maps with very high accuracy and efficiency. The comparative results demonstrate the improved accuracy of reconstructing 3D permittivity maps from the denoised C-scans rather than the noisy C-scans. The test results on new scenarios show good generalizability and robustness of the proposed scheme.

The remainder of this paper is organized as follows. The proposed network structures and learning strategy are described in Section II. In Section III, the tests on the simulation data are conducted to validate the effectiveness of the proposed method. Section IV presents generalizability and robustness tests on new scenarios. Section V shows the tests on real measurement data. Discussions and conclusions are provided in Section VI. It should be noted that, to distinguish scalars, vectors, and matrices in the rest of this paper, we use boldface lowercase letters for one-dimensional arrays (vectors) (e.g., **a**), boldface capital letters for two-dimensional arrays (matrices) (e.g., $\mathbf{A}$), boldface Euclid script letters for multi-dimensional arrays (tensors) (e.g., $\mathcal{A}$), italic lowercase letters for the scalars (e.g., $a$), italic capital letters for the tensors' dimensions (e.g., $A$), and boldface italic capital letters for the operators (e.g., $\boldsymbol{A}(\cdot)$).

## II. METHODOLOGY

In a GPR setting, a GPR system transmits EM waves into the ground and receives the signals reflected from the subsurface objects. After the GPR system is moved along the parallel traces on the $xy$ plane as demonstrated in Fig. 1, a 3D C-scan is obtained. If $\widetilde{\mathcal{X}}$ is the 3D permittivity map of the subsurface scenario, $\boldsymbol{H}(\cdot)$ represents forward modelling the scattering mechanism in the subsurface scenario, and $\mathcal{Y}$ is the resultant C-scan, then the forward problem can be described as $\mathcal{Y} = \boldsymbol{H}(\widetilde{\mathcal{X}})$. The inverse problem targets at reconstructing subsurface scenarios from C-scans, and is described as

$$\widetilde{\mathcal{X}} = \boldsymbol{H}^{-1}(\mathcal{Y}). \quad (1)$$

It converts the EM information stored in the C-scan into a 3D permittivity map of the subsurface targets, which is the subset of $\widetilde{\mathcal{X}}$ and represented as $\mathcal{X}$ in our case. The aim to solving the inverse problem based on deep learning is to find an optimal inverse mapping relationship $\boldsymbol{H}^{-1}(\cdot)$ between $\mathcal{Y}$ and $\mathcal{X}$, which can be described by a 3D DNN model, and then translate $\mathcal{Y}$ into $\mathcal{X}$ via $\boldsymbol{H}^{-1}(\cdot)$. To overcome the interference of environmental noise, we propose a two-stage scheme, named 3DInvNet, for 3D GPR data inversion. It includes a 3D denoising network (Denoiser) to obtain the denoised C-scan $\mathcal{Y}_D$ from the input noisy C-scan $\mathcal{Y}$ and a 3D inversion network (Inverter) to reconstruct $\mathcal{X}$ from $\mathcal{Y}_D$.

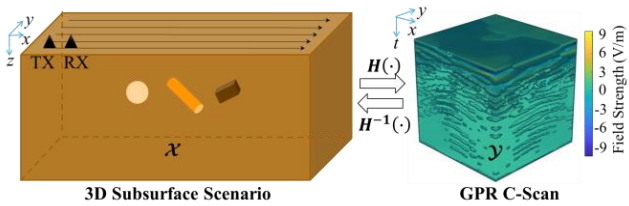

Fig. 1. Task definition of 3D GPR data inversion.

### A. Denoiser

One limitation of applying the 2D DNNs for GPR B-scan denoising in [41]-[43] to 3D cases is the high computational burden in 3D C-scan image processing. To alleviate this problem, a small 3D CNN with the feature attention mechanism [45], named Denoiser, is employed to denoise the C-scans under heterogeneous soil conditions. As shown in Fig. 2(a), the Denoiser consists of three stages: i) initial feature extraction, ii) feature learning, and iii) reconstruction. First, the initial feature extraction module, including one 3×3×3 convolutional layer with $C_1$ channels and strides of 1×1×1, captures initial features $\mathcal{F}_0 \in \mathbb{R}^{C_1 \times D \times H \times W}$ from input noisy C-scans $\mathcal{Y} \in \mathbb{R}^{D \times H \times W}$ via

$$\mathcal{F}_0 = \boldsymbol{\delta}(\boldsymbol{K}(\mathcal{Y})), \quad (2)$$

where $D$, $H$, and $W$ are spatial dimensions along principal axes. $\boldsymbol{K}(\cdot)$ represents the 3D convolutional layer. $\boldsymbol{\delta}(\cdot)$ denotes a rectified linear unit (ReLU) activation function to ensure high non-linearity of the network.

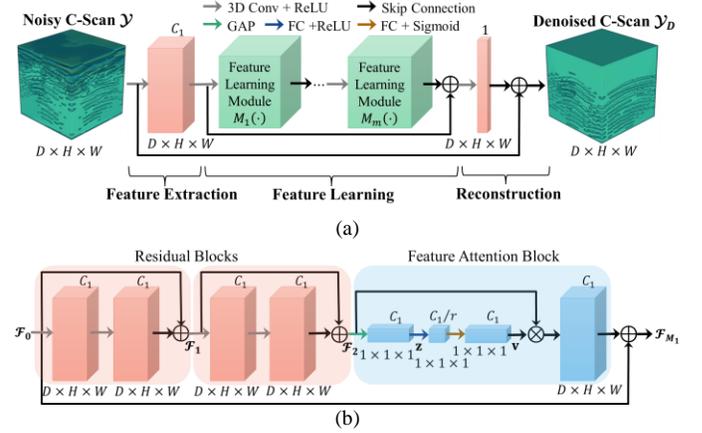

Fig. 2. (a) The network structure of Denoiser. (b) The structure of the feature learning module $\boldsymbol{M}_1(\cdot)$. "Conv", "GAP", and "FC" represent the convolutional layer, global average pooling, and fully connected layer, respectively.

Next, $m$ feature learning modules $\boldsymbol{M}(\cdot)$ are used to exploit the critical content of the extracted initial features $\mathcal{F}_0$. As shown in Fig. 2(b), every feature learning module comprises two residual blocks and one feature attention block. Each residual block has two 3×3×3 convolutional layers with identity mapping to avoid exploding gradients [46]. The operations of the residual blocks are expressed as

$$\mathcal{F}_1 = \boldsymbol{\delta}\left(\boldsymbol{K}\left(\boldsymbol{\delta}(\boldsymbol{K}(\mathcal{F}_0))\right) + \mathcal{F}_0\right) \quad (3)$$

$$\mathcal{F}_2 = \boldsymbol{\delta}\left(\boldsymbol{K}\left(\boldsymbol{\delta}(\boldsymbol{K}(\mathcal{F}_1))\right) + \mathcal{F}_1\right). \quad (4)$$

After the residual learning, a feature attention block is designed to enhance the weights of important features in $\mathcal{F}_2 \in \mathbb{R}^{C_1 \times D \times H \times W}$. In particular, the global average pooling is first applied to produce channel-wise statistics $\mathbf{z} \in \mathbb{R}^{C_1 \times 1 \times 1 \times 1}$ via

$$\mathbf{z}(c) = \frac{1}{D \times H \times W} \sum_{d=1}^{D} \sum_{h=1}^{H} \sum_{w=1}^{W} \mathcal{F}_{2_{c,d,h,w}}, \quad (5)$$

where $c$ is the channel index, and $d$, $h$, and $w$ are voxel indices along principal axes. To capture the channel-wise dependencies, a simple gating mechanism [47] that consists of two fully connected layers $\boldsymbol{F}_{\mathbf{W}}(\cdot)$ with Sigmoid function $\boldsymbol{\sigma}(\cdot)$ is adopted. The attention vector $\mathbf{v} \in \mathbb{R}^{C_1 \times 1 \times 1 \times 1}$ is computed by

$$\mathbf{v} = \boldsymbol{\sigma}\left(\boldsymbol{F}_{\mathbf{W}_1}\left(\boldsymbol{\delta}\left(\boldsymbol{F}_{\mathbf{W}_2}(\mathbf{z})\right)\right)\right), \quad (6)$$

where $\mathbf{W}_1 \in \mathbb{R}^{C_1 \times C_1/r}$ and $\mathbf{W}_2 \in \mathbb{R}^{C_1/r \times C_1}$ are weight matrices, and $r$ is a reduction ratio for limiting model complexity. Then the attended feature map is obtained by rescaling $\mathcal{F}_2 \in \mathbb{R}^{C_1 \times D \times H \times W}$ with $\mathbf{v} \in \mathbb{R}^{C_1 \times 1 \times 1 \times 1}$ via channel-wise multiplication and has the same dimensions as $\mathcal{F}_2$. With the residual connection, the output feature map $\mathcal{F}_{M_1} \in \mathbb{R}^{C_1 \times D \times H \times W}$ of the first feature learning module $\boldsymbol{M}_1(\cdot)$ is computed by

$$\mathcal{F}_{M_1} = \mathbf{v} \times \mathcal{F}_2 + \mathcal{F}_0. \quad (7)$$

As shown in Fig. 2(a), the final learned features $\mathcal{F}_{M_m} \in \mathbb{R}^{C_1 \times D \times H \times W}$ are given by

$$\mathcal{F}_{M_m} = M_m\left(M_{m-1}\left(\cdots\left(M_1(\mathcal{F}_0)\right)\right)\right). \tag{8}$$

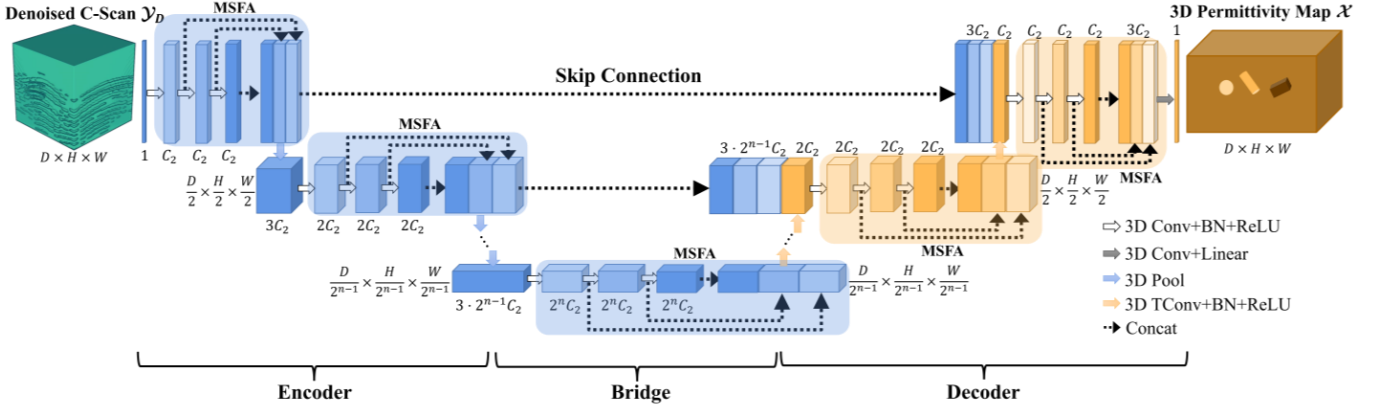

Fig. 3. The network structure of Inverter. "Conv", "BN", "ReLU", "Linear", "Pool", "TConv", "Concat", and "MSFA" represent the convolutional layer, batch normalization, rectified linear unit activation, linear activation, max-pooling layer, transposed convolutional layer, concatenation operation, and multi-scale feature aggregation module, respectively.

Finally, as shown in Fig. 2(a), a reconstruction module that includes a one-channel convolutional layer with residual learning is used to reconstruct the denoised C-scan $\mathcal{Y}_D \in \mathbb{R}^{D \times H \times W}$ from the learned feature representations via

$$\mathcal{Y}_D = \delta\left(\mathcal{Y} + K(\mathcal{F}_0 + \mathcal{F}_{M_m})\right). \tag{9}$$

### B. Inverter

After the prior denoising, a 3D inversion network, named Inverter, is designed to reconstruct the 3D subsurface permittivity map from the denoised C-scan. As shown in Fig. 3, based on 3D U-Net [44], the structure of the Inverter comprises the encoder and decoder with skip connections. In the conventional structure [44], each encoding or decoding block includes two successive convolutional layers with single scales. To extract features of subsurface objects at different scales, a multi-scale feature aggregation (MSFA) mechanism is proposed in each encoding and decoding block. Every MSFA module has three 3×3×3 convolutional layers with strides of 1×1×1. The increased number of convolutional layers aims to: 1) increase the network's depth to enhance its representation capability for the non-linear mapping from C-scans to 3D permittivity maps and 2) extract larger-scale features of objects' reflections. The receptive field (RF) size of the output feature map $\mathcal{F}_{r_f}$ of the $f$ th convolutional layer in the MSFA module is computed by [48]

$$r_f = r_{f-1} + (k_f - 1) \times \prod_{i=1}^{f-1} s_i. \tag{10}$$

$k$ and $s$ are the kernel size and stride size, which are fixed as 3 and 1, respectively. Then different RF sizes result in multiple scales [48]. As shown in Fig. 3, when the channel number is $C_2$, we concatenate $\mathcal{F}_{r_1} \in \mathbb{R}^{C_2 \times D \times H \times W}$ with the RF size of $r_1 \times r_1 \times r_1$, $\mathcal{F}_{r_2} \in \mathbb{R}^{C_2 \times D \times H \times W}$ with the RF size of $r_2 \times r_2 \times r_2$, and $\mathcal{F}_{r_3} \in \mathbb{R}^{C_2 \times D \times H \times W}$ with the RF size of $r_3 \times r_3 \times r_3$ along the channel dimension in each encoding and decoding block. The fused multi-scale feature map $\mathcal{F}_{r_{1\sim3}} \in \mathbb{R}^{3C_2 \times D \times H \times W}$ is expressed as

$$\mathcal{F}_{r_{1\sim3}} = Concat(\mathcal{F}_{r_1}, \mathcal{F}_{r_2}, \mathcal{F}_{r_3}). \tag{11}$$

Unlike the introduction of a large amount of additional convolutional layers in parallel that involves highly increased computational cost in [41], the proposed MSFA module directly combines the feature maps from successive convolutional layers with different receptive fields, which efficiently captures multi-scale features of the reflection patterns in GPR C-scans due to various properties of subsurface objects.

As shown in Fig. 3, the encoder consists of $n$ encoding blocks. Each encoding block is composed of one MSFA module and a 2×2×2 max pooling layer with strides of 2×2×2. The MSFA module extracts the objects' features from input C-scans and the compressed representation of those features is obtained after down-sampling via the pooling layer. The channels of $n$ encoding blocks are set as $[C_2, 2C_2, \cdots, 2^{n-1}C_2]$. After the encoder, one MSFA module with $2^n C_2$ channels is used to bridge the encoder and decoder. The symmetrical decoder includes $n$ decoding blocks. Each decoding block includes a 2×2×2 transposed convolutional layer with strides of 2×2×2, carried out for up-sampling, and a MSFA module, used to reconstruct the permittivity maps from the compressed representations of objects' signatures. All the transposed convolutional layers and the convolutional layers in the MSFA modules are followed by batch normalization (BN) and ReLU activations. The channels are symmetrically set as $[2^{n-1}C_2, 2^{n-2}C_2, \cdots, C_2]$. Skip connections are performed by concatenating all the convolutional layers with equal resolution in the encoder, which can provide the high-resolution features and compensate for the information loss in the down-sampling process in the encoder. In the final stage, a 3×3×3 one-channel convolutional layer with strides of 1×1×1 followed by a linear activation outputs the desired 3D permittivity map.

### C. Learning Strategy

**Step 1: Pre-train Denoiser.** A large diverse set of noisy and noise-free C-scans are used to pre-train the Denoiser. The mean square error (MSE) between the predicted denoised C-scan $\mathcal{Y}_D$ and the ground truth $\widehat{\mathcal{Y}}_D$ is used as the loss function:

$$L_1(\mathcal{Y}_D, \widehat{\mathcal{Y}}_D) = \frac{1}{D \cdot H \cdot W} \sum_{d=1}^{D} \sum_{h=1}^{H} \sum_{w=1}^{W} \left(\mathcal{Y}_{D_{d,h,w}} - \widehat{\mathcal{Y}}_{D_{d,h,w}}\right)^2. \quad (12)$$

The Adam optimizer [49] is employed to update the network parameters for minimizing the denoising loss.

**Step 2: Pre-train Inverter.** The ground truth of noise-free C-scan $\widehat{\mathcal{Y}}_D$ is used as input data of the Inverter in the pre-training phase. With the guidance of noise-free C-scans, the Inverter can accurately capture object's reflection signatures while overcoming the interference of noise patterns due to reflections from the heterogeneous soil. The mean absolute error (MAE) between the predicted permittivity map $\mathcal{X}$ and the ground truth $\widehat{\mathcal{X}}$ is used as the loss function:

$$L_2(\mathcal{X}, \widehat{\mathcal{X}}) = \frac{1}{D \cdot H \cdot W} \sum_{d=1}^{D} \sum_{h=1}^{H} \sum_{w=1}^{W} |\mathcal{X}_{d,h,w} - \widehat{\mathcal{X}}_{d,h,w}|. \quad (13)$$

The training of Inverter minimizes the inversion loss via Adam optimizer. Then in the testing phase, the denoised C-scan $\mathcal{Y}_D$ obtained from the Denoiser is fed into Inverter and the corresponding 3D permittivity map $\mathcal{X}$ is reconstructed.

**Step 3: Fine-tune the pre-trained networks.** To enhance the networks' generalizability for new scenarios that are vastly different from the pre-training dataset, transfer learning [50] is further employed via fine-tuning the pre-trained networks obtained in steps 1-2. First, a small additional set of data with new scenarios are generated. Second, the pre-trained networks are set as the initial states and the network parameters are further updated based on the loss functions (12) and (13) using the new training dataset until convergence. Finally, the fine-tuned networks with the transferred data domain can predict the denoised C-scans and reconstruct 3D permittivity maps for the new scenarios.

## III. NUMERICAL EXPERIMENTS

### A. Dataset Generation

*Dataset I.* To train and test the proposed networks, a dataset covering diverse subsurface scenarios is generated. Each data set includes a noisy C-scan $\mathcal{Y}$, a noise-free C-scan $\widehat{\mathcal{Y}}_d$, and a corresponding 3D permittivity map $\widehat{\mathcal{X}}$. The numerical dataset is generated by an open-source 3D finite-difference time-domain simulator, gprMax [51]-[52]. As shown in Fig. 1, a 1×1×0.26 m³ soil environment is considered. The resolution is 2.5×2.5×2.5 mm³. The relative permittivity and conductivity of the soil are set as 4 and 0, respectively. A Hertzian dipole transmits a Ricker signal with a center frequency of 1 GHz. The time window is set to 15 ns. A probe positioned 10 cm away from the dipole is used to collect the reflected signals. Both the dipole and probe are 2 cm above the ground. The scanning trajectories are shown as the black lines in Fig. 1. There are 12 scanning lines with ten evenly spaced scanning points on every line. The shapes of subsurface objects are randomly selected as cylinder, sphere, or box. The permittivity of each object is selected in the range of [8, 27]. For the cylinder, the radius and length are randomly selected from the intervals of [0.02, 0.05] m ([0.29, 0.73] wavelength) and [0.01, 0.33] m ([0.15, 4.84] wavelength), respectively. For the sphere, the radius is randomly selected from [0.02, 0.05] m ([0.29, 0.73] wavelength). For the box, all the edge lengths are randomly selected from [0.04, 0.1] m ([0.59, 1.47] wavelength). It should be noted that the wavelength is calculated based on the highest significant frequency of 2.2 GHz. The subsurface object is allowed to be positioned in a domain of 0.4×0.4×0.26 m³ centered in the middle of the soil to guarantee complete hyperbolic patterns. The simulated C-scans are processed by time-zero correction and mean subtraction [32] to remove direct coupling signals and reflections from the ground, and then used as ground-truth noise-free C-scans $\widehat{\mathcal{Y}}_D$.

To model environmental noise due to the reflections from the heterogeneous soil, 3D Peplinski mixing models [53]-[54] are used to build soil environments without any objects. The soil properties are set as sand fraction 0.6, clay fraction 0.4, bulk density 2 g/cm³, and sand particle density of 2.66 g/cm³. 50 different materials described by 50 Debye functions over a range of water volumetric fractions from 0.1% to 10% are randomly distributed in the soil, producing a relative permittivity range of [3.63, 7.42] and a conductivity range of [0.01, 0.05] S/m. To increase the diversity of noise patterns, ten random heterogeneous distributions are generated, and the simulated C-scans contain only environmental noise. After time-zero correction and mean subtraction [32], the noise-only C-scan is added to the noise-free C-scan $\widehat{\mathcal{Y}}_D$ to form a synthetic noisy C-scan $\mathcal{Y}$. Finally, 5850 sets of [$\mathcal{Y}, \widehat{\mathcal{Y}}_D, \widehat{\mathcal{X}}$] (1950 one-object scenarios and 3900 two-object scenarios) are produced for pre-training and 150 (50 one-object scenarios and 100 two-object scenarios) are generated for testing.

*Dataset II.* To further apply the proposed method to heterogeneous scenarios, rather than the direct addition in dataset I, the objects are buried in the heterogeneous soil and the GPR scanning on the $xy$ plane is conducted to obtain the noisy C-scan $\mathcal{Y}$. It should be noted that the random distribution of the heterogeneous soil is new as well. The ground-truth noise-free C-scan $\widehat{\mathcal{Y}}_D$ is obtained via subtracting the noise-only C-scan from the noisy C-scan. The background permittivity in the ground-truth permittivity maps is set to 5.72, which is the average relative permittivity of the 50 materials in the heterogeneous soil model. With randomly selected properties of subsurface objects, 285 sets of [$\mathcal{Y}, \widehat{\mathcal{Y}}_D, \widehat{\mathcal{X}}$] (95 one-object scenarios and 190 two-object scenarios) are generated to fine-tune the pre-trained Denoiser and Inverter, and 15 (5 one-object scenarios and 10 two-object scenarios) are generated for testing.

### B. Implementation Details

The proposed scheme is implemented on PyTorch [55] and executed on an NVIDIA GeForce RTX 3090 GPU. Every C-scan is normalized to the range of [0, 1]. Both the C-scans and permittivity maps are resized to 128×128×128. The noisy C-scan $\mathcal{Y}$ and noise-free C-scans $\widehat{\mathcal{Y}}_D$ in dataset I are used for pre-training Denoiser based on (12). The noise-free C-scans $\widehat{\mathcal{Y}}_D$ and 3D permittivity maps $\widehat{\mathcal{X}}$ in dataset I are employed for pre-training Inverter based on (13). $m$ and $C_1$ in the Denoiser are set as 2 and 8, respectively. $n$ and $C_2$ in the Inverter are set as 4 and 8, respectively. The learning rate is initially 0.001 and decreased to 98% of its value if the training loss has no drop in the last epoch. The Denoiser and Inverter are separately pre-trained for 100 epochs. The models with the lowest validation losses are used to test the performance. In the testing stage, the noisy C-scan $\mathcal{Y}$ is fed into the pre-trained Denoiser, then the denoised C-scan $\mathcal{Y}_D$ is input into the pre-trained Inverter, which finally outputs the 3D permittivity map $\mathcal{X}$. Using the pre-trained Denoiser and Inverter as the initial state, the networks are further fine-tuned with dataset II. To avoid over-fitting, the learning rate is set as small as 0.0006 and decreased to 99% of its value if the training loss has no drop in the last epoch.

The structural similarity (SSIM), peak signal-to-noise ratio (PSNR), MAE, and mean relative error (MRE) are defined to analyze the denoising performance of Denoiser. The SSIM, MSE, MAE, and mean absolute percentage error (MAPE) are used to evaluate the inversion performance of Inverter. These metrics are expressed as:

$$SSIM(\mathcal{T},\mathcal{P}) = \frac{(2\mu_\mathcal{P}\mu_\mathcal{T}+c_1)(2\sigma_{\mathcal{PT}}+c_2)}{(\mu_\mathcal{P}^2+\mu_\mathcal{T}^2+c_1)(\sigma_\mathcal{P}^2+\sigma_\mathcal{T}^2+c_2)}, \quad (14)$$

$$MSE(\mathcal{T},\mathcal{P}) = \frac{1}{D\cdot H\cdot W}\sum_{d=1}^{D}\sum_{h=1}^{H}\sum_{w=1}^{W}(\mathcal{P}_{d,h,w}-\mathcal{T}_{d,h,w})^2, \quad (15)$$

$$PSNR = 10\log_{10}\frac{1}{MSE(\mathcal{T},\mathcal{P})}, \quad (16)$$

$$MAE(\mathcal{T},\mathcal{P}) = \frac{1}{D\cdot H\cdot W}\sum_{d=1}^{D}\sum_{h=1}^{H}\sum_{w=1}^{W}|\mathcal{P}_{d,h,w}-\mathcal{T}_{d,h,w}|, \quad (17)$$

$$MRE(\mathcal{T},\mathcal{P}) = \frac{\sqrt{\sum_{d=1}^{D}\sum_{h=1}^{H}\sum_{w=1}^{W}(\mathcal{P}_{d,h,w}-\mathcal{T}_{d,h,w})^2}}{\sqrt{\sum_{d=1}^{D}\sum_{h=1}^{H}\sum_{w=1}^{W}\mathcal{T}_{d,h,w}^2}}\cdot 100\%, \quad (18)$$

$$MAPE(\mathcal{T},\mathcal{P}) = \frac{1}{D\cdot H\cdot W}\sum_{d=1}^{D}\sum_{h=1}^{H}\sum_{w=1}^{W}\frac{|\mathcal{P}_{d,h,w}-\mathcal{T}_{d,h,w}|}{|\mathcal{T}_{d,h,w}|}\cdot 100\%. \quad (19)$$

In these equations, $\mathcal{P}$ represents the predicted result, which can be the denoised C-scan $\mathcal{Y}_D$ or the reconstructed 3D permittivity map $\mathcal{X}$. $\mathcal{T}$ represents the ground truth $\widehat{\mathcal{Y}}_D$ or $\widehat{\mathcal{X}}$. $\mu_\mathcal{P}$ and $\mu_\mathcal{T}$ are means of $\mathcal{P}$ and $\mathcal{T}$, respectively. $\sigma_\mathcal{P}$ and $\sigma_\mathcal{T}$ are variances of $\mathcal{P}$ and $\mathcal{T}$, respectively. $\sigma_{\mathcal{PT}}$ is the covariance of $\mathcal{P}$ and $\mathcal{T}$. $c_1 = (0.01i)^2$ and $c_2 = (0.03i)^2$ are two variables where $i$ is the dynamic range of image pixels. The average of each metrics of all testing samples is calculated to evaluate the performance.

### C. Result Analysis

*1) Quantitative Results:* The evaluation metrics of the 3D denoising performance are listed in Table I. For the testing data with additional noise in dataset I, the well-trained Denoiser achieves high SSIM and PSNR and low MAE and MRE. For the simulated noisy data in dataset II, using the fine-tuned network, the denoising accuracy is still high. Even though the reflection patterns in C-scans of dataset II is different from those in dataset I, the transfer learning can adapt the network pre-trained by dataset I to the new dataset II, which effectively enhances the generalizability of the network.

TABLE I
DENOISING EVALUATION METRICS ON DATASETS I AND II

| Dataset | # of Testing Samples | SSIM (×10⁻²) (↑) | PSNR (dB) (↑) | MAE (×10⁻⁴) (↓) | MRE (%) (↓) |
|---|---|---|---|---|---|
| I | 150 | 99.96 | 62.99 | 3.78 | 0.14 |
| II | 15 | 99.98 | 66.95 | 2.08 | 0.09 |

TABLE II
INVERSION EVALUATION METRICS ON DATASETS I AND II

| Groups | Amount | | SSIM (×10⁻²) (↑) | | MSE (×10⁻²) (↓) | | MAE (×10⁻²) (↓) | | MAPE (%) (↓) | |
|---|---|---|---|---|---|---|---|---|---|---|
| | I | II | I | II | I | II | I | II | I | II |
| i | 50 | 5 | 99.91 | 99.65 | 3.04 | 1.81 | 0.49 | 0.76 | 0.07 | 0.11 |
| ii | 67 | 6 | 99.73 | 99.26 | 8.90 | 8.06 | 1.10 | 1.64 | 0.15 | 0.20 |
| iii | 33 | 4 | 99.63 | 99.10 | 15.12 | 19.30 | 1.74 | 2.89 | 0.21 | 0.31 |
| i, ii, and iii | 150 | 15 | 99.77 | 99.34 | 8.31 | 8.98 | 1.04 | 1.68 | 0.14 | 0.20 |

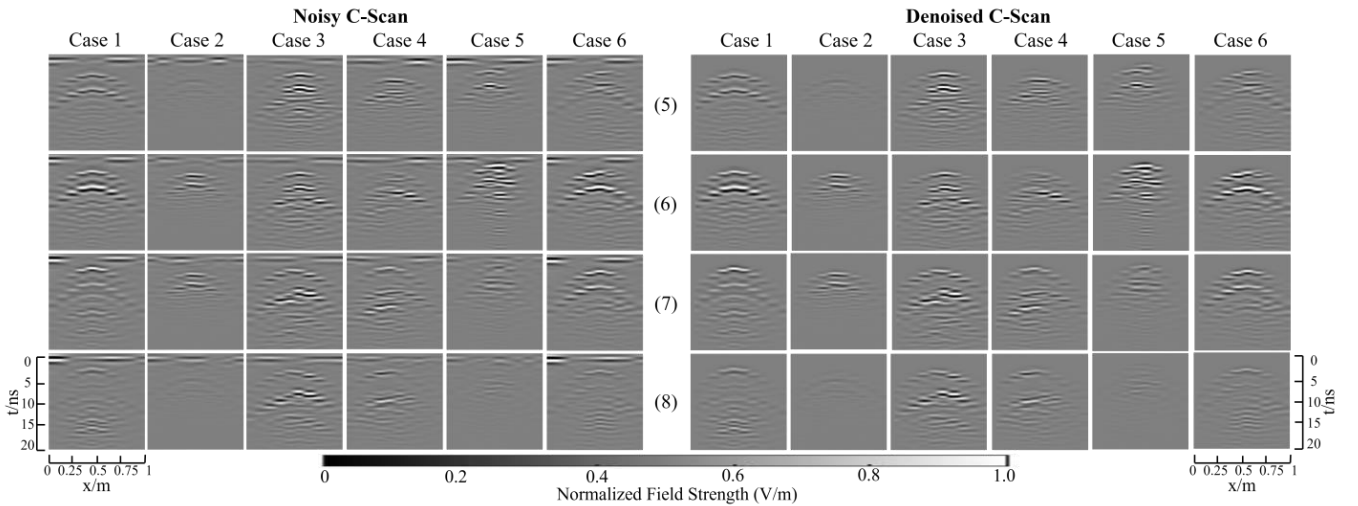

Fig. 4. The imaging results of denoised C-scans predicted by Denoiser with dataset I. It should be noted that each C-scan includes 12 B-scans and selected B-scans (6)-(8) with the main reflection signatures are shown.

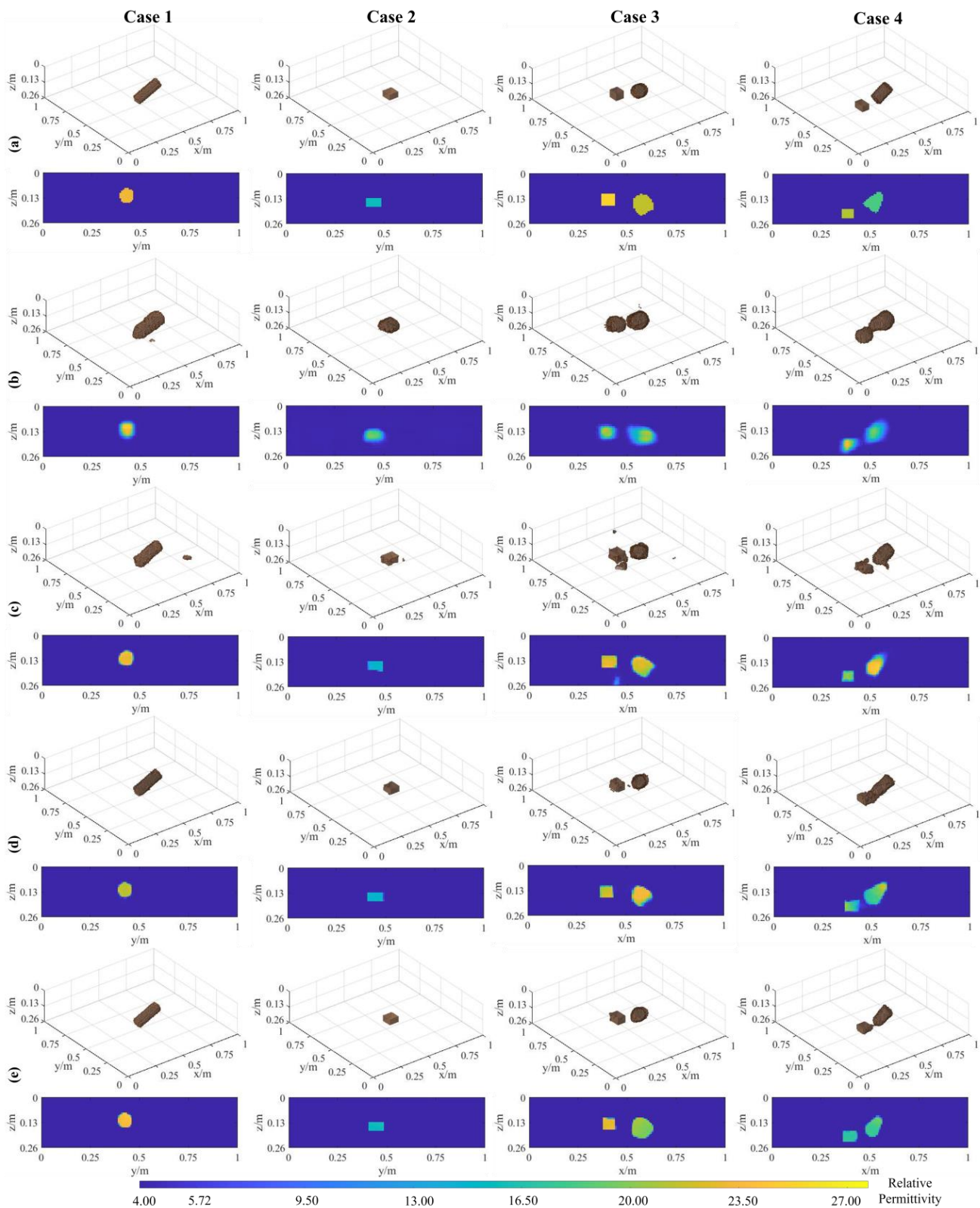

Fig. 5. The imaging results of 3D permittivity maps predicted by the proposed 3DInvNet for one-object scenarios (Cases 1-2) and two-separated-object scenarios (Cases 3-4) in dataset I. While (a) shows the ground truths, (b)-(e) show the predictions using Models (1)-(4), respectively.

To quantitatively analyze the accuracy of 3D permittivity map reconstruction using the Inverter, we classify the testing data into three groups: i) one-object cases, ii) two-separated-object cases, and iii) two-overlapping-object cases. The metrics on the three types of data in datasets I and II are shown in Table II. Clearly, for both datasets I and II, the SSIMs, MSEs, MAEs, and MAPEs of one-object cases (group i) are the best. For the two-separated-object cases (group ii), the SSIMs are lower and MSEs, MAEs, and MAPEs are higher than those in one-object cases. The performance degradation is because the subsurface scenarios with two objects are more complex, which makes reconstructing two-object permittivity maps much more challenging than one-object cases. The hyperbolic signatures in the C-scans in two-overlapping-object cases (group iii) are even more complicated, increasing the difficulty of GPR data inversion. This is evidenced by the decreases in SSIMs and increases in MSEs, MAEs, and MAPEs. On average, the high SSIM and low MSE, MAE, and MAPE of the total 150 testing data combining all these three cases in dataset I prove the capability of Inverter in accurately reconstructing permittivity maps. Compared to the results of dataset I, the average SSIM of the 15 testing sets in dataset II is slightly lower and the average MSE, MAE, and MAPE are higher but remain satisfactory.

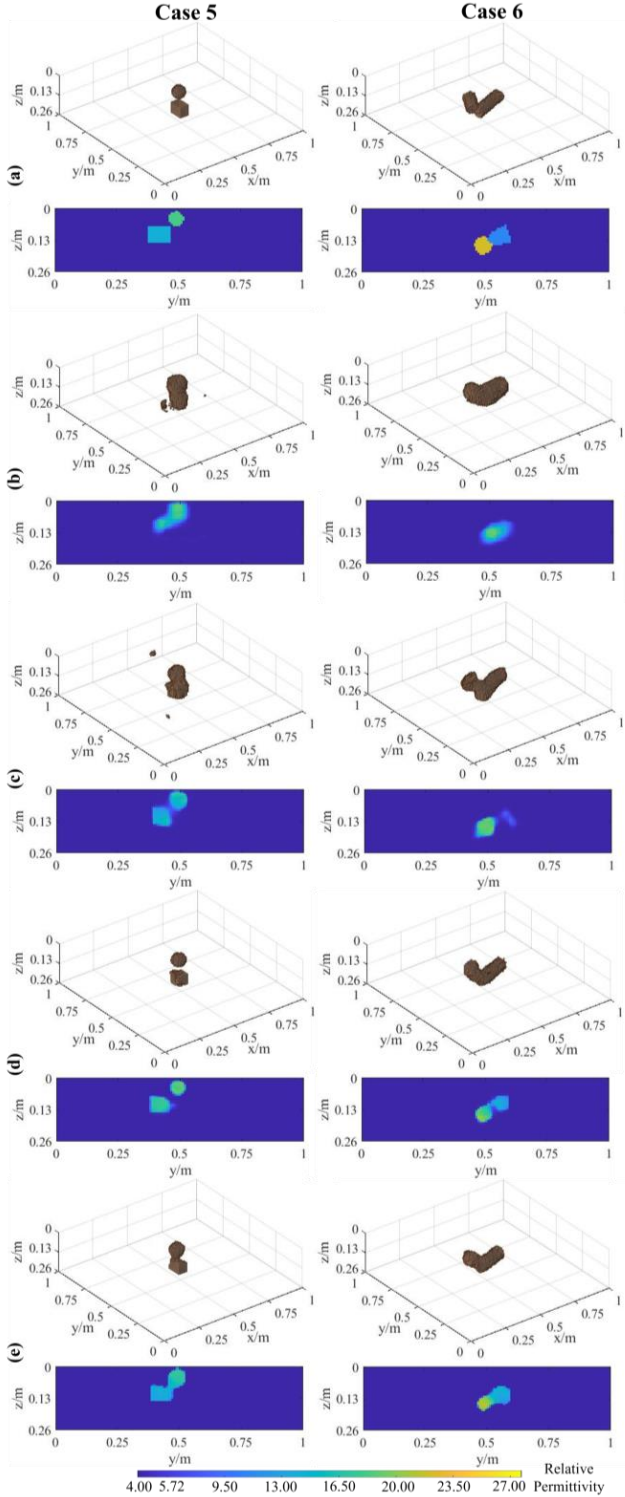

Fig. 6. The imaging results of 3D permittivity maps for two-overlapping-object scenarios (Cases 5-6) in dataset I. While (a) shows the ground truths, (b)-(e) show the predictions using Models (1)-(4), respectively.

As for the inversion efficiency, the proposed scheme outperforms the classical iterative algorithms. For example, in the most conservative scenario, one iteration of FWI requires computing a C-scan for a given permittivity map, which takes over 40 minutes when using the simulator gprMax on a GPU platform in our case. Moreover, FWI often requires more than one iteration [7]-[8], which is certainly time-consuming. However, as shown in Table III, the total testing time using the proposed 3DInvNet, including the Denoiser and Inverter, to generate one permittivity map is only 0.59 s, which is at least 4000 times less than that required by one iteration of FWI. Although the pre-training and fine-tuning are computationally expensive, they only need to be done once. Using the well-trained networks, the 3D permittivity maps can be predicted from the C-scans in near real-time.

TABLE III
COMPUTATION TIME OF THE PROPOSED 3DINVNET

| Network | # of Parameters ($\times 2^{20}$) | Pre-Training Time (h) | Fine-Tuning Time (h) | Testing Time (s) |
| --- | --- | --- | --- | --- |
| Denoiser | 0.01 | 27.50 | 2.05 | 0.20 |
| Inverter | 3.06 | 58.33 | 3.70 | 0.39 |

*2) Imaging Results:* For dataset I, the imaging results of the denoised C-scans obtained from Denoiser are shown in Fig. 4. Cases 1-2, Cases 3-4, and Cases 5-6 present the one-object scenarios (group i), two-separated-object scenarios (group ii), and two-overlapping-object scenarios (group iii), respectively. To visualize the denoised C-scans clearly, selected B-scans [(5)-(8)] that include the main reflection signatures in each C-scan are plotted. Compared with the noisy C-scans, the noise due to reflections from the heterogeneous soil environments is highly suppressed in the denoised C-scans. The reflection patterns of subsurface objects are extracted and used as the inputs of Inverter to reconstruct 3D permittivity maps.

As shown in Figs. 5-6, the 3D permittivity maps predicted by the Inverter [Fig. 5(e) and Fig. 6(e)] are compared with the ground-truth 3D permittivity maps [Fig. 5(a) and Fig. 6(a)]. Both the 3D geometry and one-slice of permittivity map are presented. In the 3D geometry, the soil environment is set as transparent, and the objects are shown in brown. Cases 1-2 show the results for one-object scenarios. The predicted geometries match well with the actual geometries in the ground truth. To further evaluate the accuracy in estimating the permittivity map, the ground truth and prediction of permittivity maps on a cross-section are compared. It is observed that the predicted permittivity map, including the value and position, is quite close to the ground truth. The imaging results of the two-separated-object cases are shown as Cases 3-4. Compared to the one-object scenarios, two objects introduce more complicated signatures in the C-scan, such as the interleaved hyperbolas and misleading multiple reflected noise. As presented, the proposed method still achieves high accuracy. Even though these objects buried in the soil have various shapes, locations, sizes, and permittivity values, both the predicted 3D geometry and one-slice permittivity map match well with the ground truths.

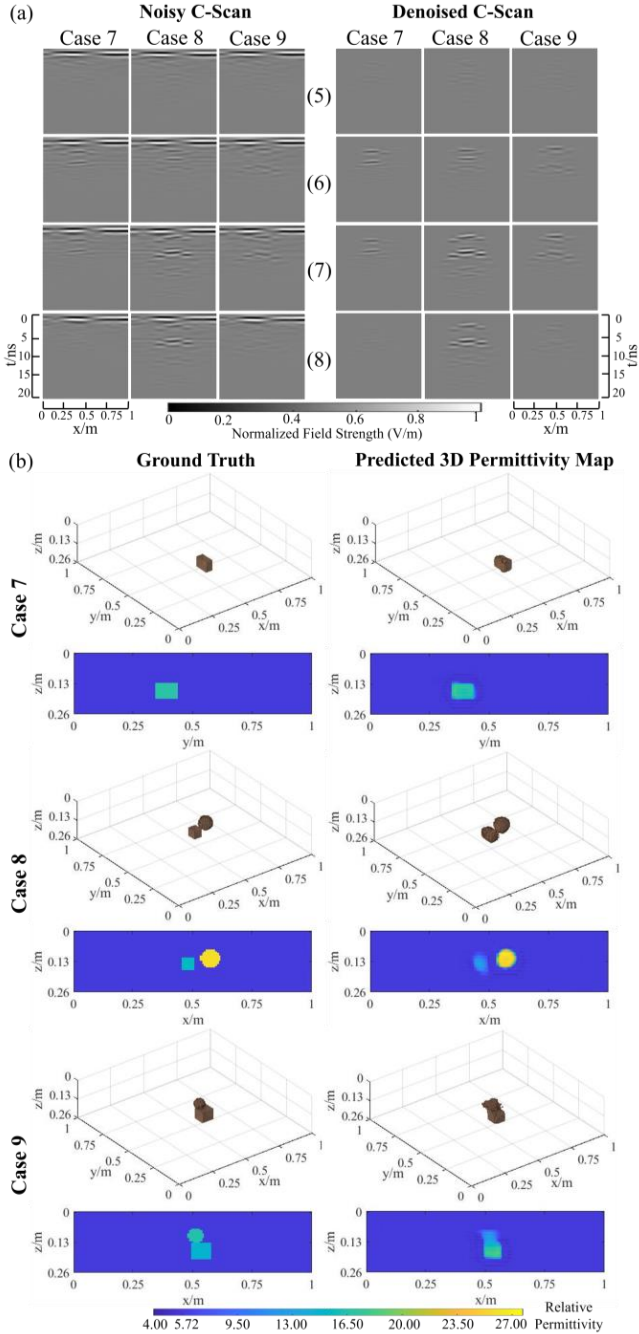

Fig. 7. (a) Denoised C-scans obtained from Denoiser for dataset II. (b) Reconstructed 3D permittivity maps obtained from Invertor for dataset II.

To verify the high performance in the resolution of our proposed scheme, the inversion results for overlapped and inhomogeneous objects are shown as Cases 5-6 in Fig. 6. Due to more complex subsurface distributions, reconstructing 3D permittivity maps from the C-scans becomes more challenging. The imaging results show that the predicted permittivity maps as shown in Fig. 6(e) are close to the ground truth ones as shown in Fig. 6(a). Even though there is some mismatch on the objects' edges, the subsurface objects can still be clearly recognized.

To further demonstrate the imaging performance for real simulated noisy data, Fig. 7 presents the denoising and inversion results for dataset II. Figs. 7(a)-(b) show the denoised C-scans and reconstructed 3D permittivity maps, respectively. Cases 7-9 provide three examples in groups i-iii, respectively. As shown, using the denoised C-scans with negligible clutters obtained from the Denoiser as inputs of the Inverter, the predicted 3D permittivity maps match the ground truths. Although the two-object cases are not as precise as the results of dataset I using the networks trained with a large set of data, the subsurface objects can still be recognized well.

### D. Comparative Study

To prove the effectiveness of the proposed Denoiser and the MSFA modules in the Inverter, the inversion metrics of four ablation models on dataset I are compared, as shown in Table IV. The four models include (1) the baseline 3D U-Net [44] with three successive convolutional layers in each encoding or decoding block, (2) only the Inverter, (3) the Denoiser and the baseline 3D U-Net [44], and (4) the proposed model. As shown in Table IV, Model (1) performs the worst due to the interference of environmental noise in the permittivity reconstruction process and the limited perception view of the network. The performance of Model (2) is improved with the improved multi-scale perceptive capability introduced by the MSFA modules, but the performance is still limited by the interference of the environmental noise in C-scans. Compared with Model (1), Model (3) shows dramatically improved reconstruction accuracy with the noise suppression process in the Denoiser. Our proposed model (Model (4)) shows that first denoising the C-scan with Denoiser and then reconstructing permittivity maps with Inverter achieves the highest accuracy.

TABLE IV
INVERSION METRICS COMPARISON FOR COMPARATIVE STUDY ON DATASET I

| Model | Denoiser | MSFA Module | SSIM (×10⁻²) (↑) | MSE (×10⁻²) (↓) | MAE (×10⁻²) (↓) | MAPE (%) (↓) |
|---|---|---|---|---|---|---|
| (1) | ✗ | ✗ | 99.06 | 14.12 | 2.72 | 0.46 |
| (2) | ✗ | ✓ | 99.19 | 9.01 | 2.75 | 0.56 |
| (3) | ✓ | ✗ | 99.55 | 9.32 | 1.92 | 0.36 |
| (4) | ✓ | ✓ | **99.77** | **8.31** | **1.04** | **0.14** |

The imaging results for one-object and two-separated-object scenarios in dataset I using the four models are compared in Figs. 5(b)-(e), respectively, and those for two-overlapping-object scenarios are shown in Figs. 6(b)-(e), respectively. It can be observed that Model (1) performs the worst as shown in Fig. 5(b) and Fig. 6(b). The shapes and permittivity values of the reconstructed objects are far from their ground truths. As shown in Fig. 5(c) and Fig. 6(c), without the Denoiser, Model (2) incorrectly identifies the noise patterns as objects' reflections and reconstructs some clutters in the predicted 3D permittivity maps. As shown in Fig. 5(d) and Fig. 6(d), the clutters in the predicted maps using Model (3) are decreased due to the introduction of Denoiser, but the permittivity values and sizes of subsurface objects are not accurate especially for two-object scenarios as shown in Cases 3-6. This is due to the limited multi-scale feature extraction capability of the inversion network. As shown in Fig. 5(e) and Fig. 6(e), the reconstructed 3D permittivity maps using Model (4) that combines the Denoiser and MSFA modules in the Inverter are the closest to the ground truths regarding the sizes, locations, shapes, orientations, and permittivity values of the objects.

### E. Hyperparameter Selection

To analyze the hyperparameter selection of the proposed scheme, the evaluation metrics of Denoiser and Inverter with different parameters are compared in Table V. As shown, for the Denoiser, a larger number of feature learning modules $m$ and a higher channel number $C_1$ lead to more accurate denoising results. For the Inverter, the reconstruction accuracy increases as the number of the encoding/decoding blocks $n$ and the channel number $C_2$ increase. These improvements are due to the ability of the deepened and widened networks to describe more accurate mapping relationships between the input and output images. However, the increase of these hyperparameters also increases the computational costs of the network training and testing, especially for the 3D case. According to the imaging results shown in Figs. 4-7, when $m$ and $C_1$ of the Denoiser reach to 2 and 8, respectively, the denoised C-scans show clear reflection signatures of subsurface objects; when $n$ and $C_2$ of the Inverter reach to 4 and 8, respectively, the reconstructed maps achieve high accuracy. Therefore, these hyperparameter values are selected to balance the imaging performance and computational burden in our case.

## IV. GENERALIZABILITY AND ROBUSTNESS TESTS

### A. Zero/Three-Object Scenarios

*Dataset III.* To evaluate the generalization capability of the proposed scheme, the pre-trained Denoiser and Inverter are employed to denoise C-scans and restore 15 scenarios with three objects or without any objects buried in the heterogeneous soil. It should be noted that only one-object and two-object scenarios are included in the pre-training dataset (dataset I), so the three-object cases and no-object cases are completely new scenarios to the denoising and inversion networks. Other properties of soil and objects and the dataset generation procedure are the same as those of dataset I.

*Test Results.* As shown in Table VI (dataset III), without any additional retraining or fine-tuning, the evaluation metrics of Denoiser and Inverter obtained for this testing dataset are still highly satisfactory. Using the denoised C-scans obtained from the Denoiser as inputs of the Inverter, the imaging results of reconstructed 3D permittivity maps are shown in Fig. 8(a). Case 10 shows the imaging result when no object is buried in the soil. Both the reconstructed geometry and one-slice permittivity map contain only the soil environment. Cases 11 and 12 are the scenarios when there are three buried objects with various random properties. It is clear from the 3D geometry and one-slice permittivity map that the predicted results match well with the ground truth. Despite the training dataset does not cover the no-object and three-object scenarios, the proposed scheme can still reconstruct their 3D permittivity maps with good accuracy.

### B. New Heterogeneous Soil Condition

*Dataset IV.* To evaluate the robustness of the proposed method in reconstructing subsurface scenarios from C-scans with new and different noise patterns, the 3D Peplinski mixing model with a new heterogeneous soil condition is constructed. Different from the soil properties in dataset II, the sand fraction, clay fraction, water content range, relative permittivity range, conductivity range are set as 0.5, 0.5, [0.1%, 15%], [3.62, 8.59], and [0.01, 0.06] S/m, respectively. The background permittivity in the ground-truth permittivity maps is set to 6.34, which is the average relative permittivity of the 50 materials in the new soil model. 285 additional training data sets are generated using the new soil model to fine-tune the pre-trained networks and 15 sets are produced for testing. The implementation details are the same as those for dataset II.

*Test Results.* As shown in Table VI (dataset IV), the high SSIM and PSNR and low MAE and MRE of the Denoiser indicate high denoising accuracy for the C-scans obtained under the new heterogeneous soil environment. Then the subsequent

TABLE V
EVALUATION METRICS ON DATASET I USING DIFFERENT HYPERPARAMETERS

| Denoiser | | SSIM ($\times 10^{-2}$) | PSNR (dB) | MAE ($\times 10^{-4}$) | MRE (%) |
| --- | --- | --- | --- | --- | --- |
| $m$ | $C_1$ | ($\uparrow$) | ($\uparrow$) | ($\downarrow$) | ($\downarrow$) |
| 0 | 8 | 99.63 | 53.04 | 9.43 | 0.45 |
| 1 | 8 | 99.88 | 57.86 | 6.02 | 0.26 |
| **2** | **8** | **99.96** | **62.99** | **3.78** | **0.14** |
| 2 | 4 | 99.81 | 55.67 | 7.88 | 0.34 |
| 2 | 2 | 98.75 | 46.57 | 13.86 | 0.95 |
| Inverter | | SSIM ($\times 10^{-2}$) | MSE ($\times 10^{-2}$) | MAE ($\times 10^{-2}$) | MAPE (%) |
| $n$ | $C_2$ | ($\uparrow$) | ($\downarrow$) | ($\downarrow$) | ($\downarrow$) |
| 2 | 8 | 98.23 | 10.67 | 3.67 | 0.75 |
| 3 | 8 | 98.93 | 7.77 | 2.96 | 0.61 |
| **4** | **8** | **99.77** | **8.31** | **1.04** | **0.14** |
| 4 | 4 | 99.63 | 8.50 | 1.54 | 0.24 |
| 4 | 2 | 99.65 | 9.77 | 1.65 | 0.26 |

TABLE VI
EVALUATION METRICS OF GENERALIZATION AND ROBUSTNESS TESTS

| Denoiser | SSIM ($\times 10^{-2}$) ($\uparrow$) | PSNR (dB) ($\uparrow$) | MAE ($\times 10^{-4}$) ($\downarrow$) | MRE (%) ($\downarrow$) |
| --- | --- | --- | --- | --- |
| III | 99.71 | 68.52 | 3.73 | 0.29 |
| IV | 99.99 | 69.97 | 1.41 | 0.07 |
| V | 99.97 | 60.26 | 5.15 | 0.20 |
| Inverter | SSIM ($\times 10^{-2}$) ($\uparrow$) | MSE ($\times 10^{-2}$) ($\downarrow$) | MAE ($\times 10^{-2}$) ($\downarrow$) | MAPE (%) ($\downarrow$) |
| III | 99.78 | 13.13 | 1.35 | 0.16 |
| IV | 98.97 | 13.10 | 2.82 | 0.35 |
| V | 98.68 | 9.33 | 3.31 | 0.66 |

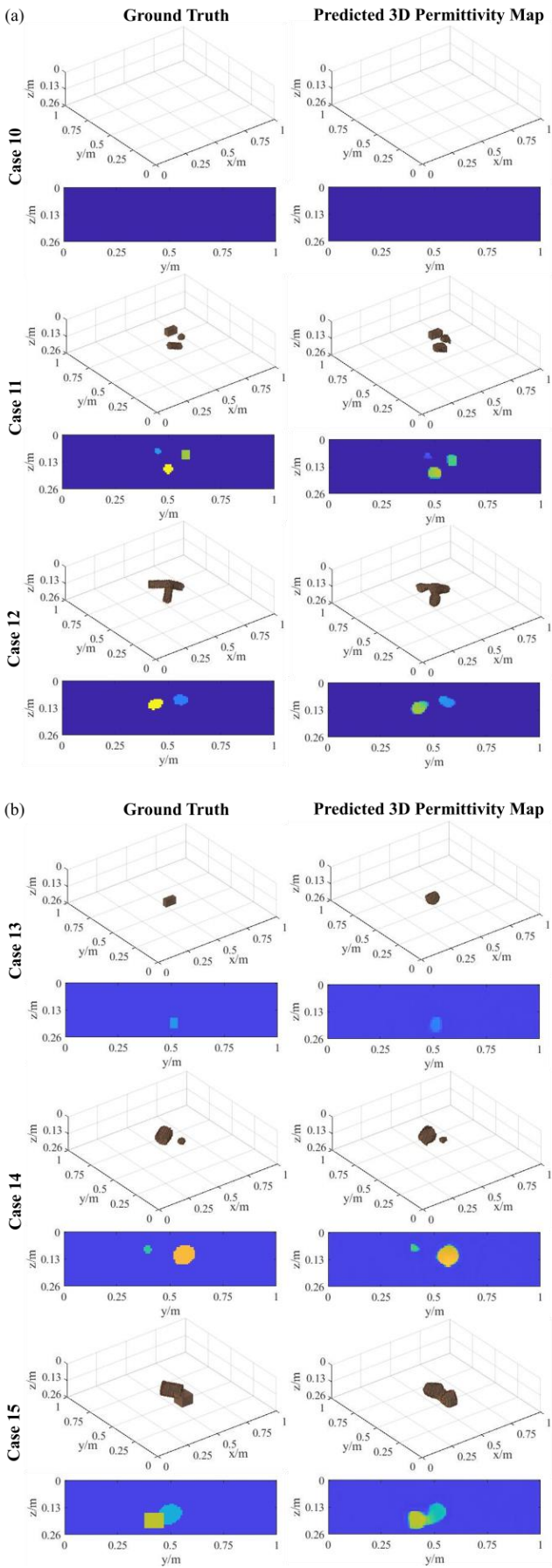
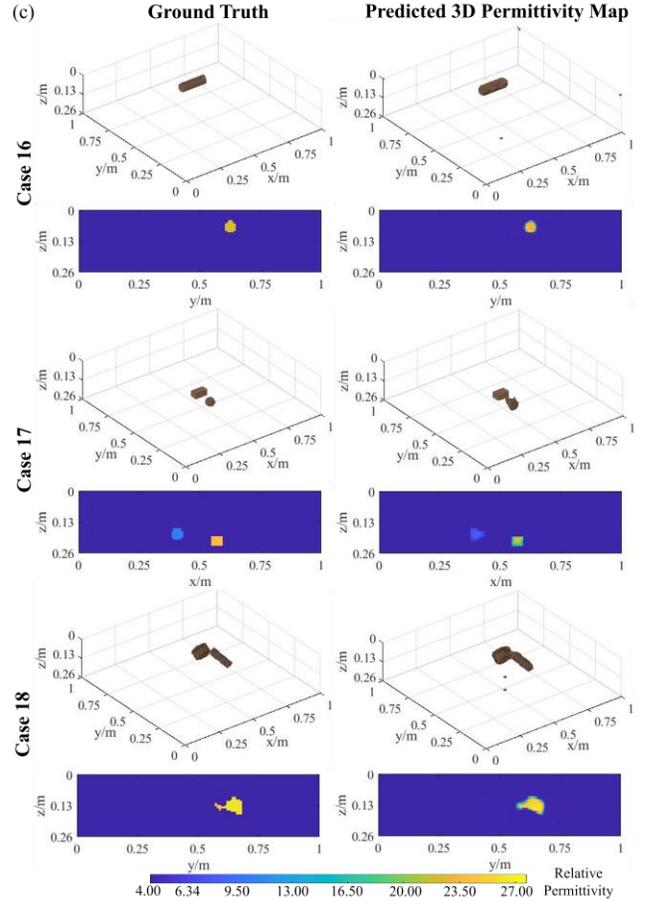

Fig. 8. The imaging results of generalizability and robustness tests on (a) dataset III, (b) dataset IV, and (c) dataset V.

Inverter reconstructs 3D permittivity maps from denoised C-scans in high SSIM and low MSE, MAE, and MAPE. Cases 13-15 in Fig. 8(b) show three imaging examples of the inversion results, including one-object, two-separated-object, and two-overlapping-object scenarios. It is clear from these visualized results that the predicted 3D permittivity maps are close to the ground truths, and the soil's permittivity and subsurface objects' properties are restored accurately.

### C. Commercial GPR Antenna Model

*Dataset V.* To further test the generalization capability of the proposed scheme on a realistic commercial antenna, the GSSI 400-MHz GPR antenna model [56] is used in gprMax to transmit and receive the signals. Compared to the datasets I-IV obtained from the dipole/probe configuration at 1 GHz, both the operating frequency and the TX/RX configurations are the ones changed, and the height between the antenna model and ground is set to zero, resulting in highly different reflection patterns in C-scans. Other properties of subsurface scenarios are kept the same as those in dataset I. 285 data sets obtained with the GSSI 400-MHz GPR antenna model are produced to fine-tune the pre-trained networks. The fine-tuned networks are used to denoise 15 noisy C-scans and predict 3D permittivity maps.

*Test Results.* As shown in Table VI (dataset V), the satisfactory denoising performance with the commercial GSSI 400-MHz antenna model is demonstrated by the high SSIM and PSNR and low MAE and MRE of the Denoiser. As for the inversion performance, the high SSIM and low MSE, MAE, and MAPE of the Inverter indicate good accuracy of 3D permittivity map reconstruction. To visually analyze the inversion results, the predicted 3D permittivity maps with the ground truths of the one-object, two-separated-object, and two-overlapping-object scenarios are compared, as shown in Cases 16-18 in Fig. 8(c). It is clear from the imaging results that the predicted permittivity maps match well with their corresponding ground truths. The characteristics of the subsurface objects are restored with high accuracy. These results verify that by using a small additional dataset to fine-tune the pre-trained networks, the proposed 3DInvNet yields an excellent generalization capability on a commercial antenna system with new operating frequencies and configurations.

## V. Tests with Real Measurement Data

### A. Dataset Collection and Experimental Setup

GPR experiments are carried out in a sandy test field to verify the applicability and effectiveness of the proposed scheme. As shown in Fig. 9(a), the commercial GSSI's Utility Scan Pro GPR system with a 400-MHz antenna is used to obtain B-scans and then stack them into C-scans. The scanning area is 1×1 m$^2$. Every C-scan includes 21 B-scans and the distance between two adjacent traces is 5 cm. Every B-scan consists of 88 A-scans and the number of sampling points in every A-scan is 512. The time window is set as 20 ns. As shown in Fig. 9(b), the scanning area is rotated to model various horizontal orientations of objects and shifted along the $x$ and $y$ axes to realize the scanning of the objects from various positions. Other properties, such as depth and vertical orientation, are changed via burying the objects. The buried objects are selected from five wooden objects with various shapes, sizes, and permittivity values. The relative permittivities ($\varepsilon_r$) of the objects and the sand are measured by Keysight N1501A Dielectric Probe as shown in Fig. 10(a). The average value of relative permittivities of five sand samples is 3.65, which is set as the background relative permittivity in the ground truths for training the proposed network. The properties of the buried objects in the experiments are shown in Fig. 10(b). The ranges of the $x$-coordinate, $y$-coordinate, $z$-coordinate (depth), horizontal angle, and vertical angle of the objects are [35, 65] cm, [35, 65] cm, [9, 25] cm, [0, 60] degree, and [0, 60] degree, respectively.

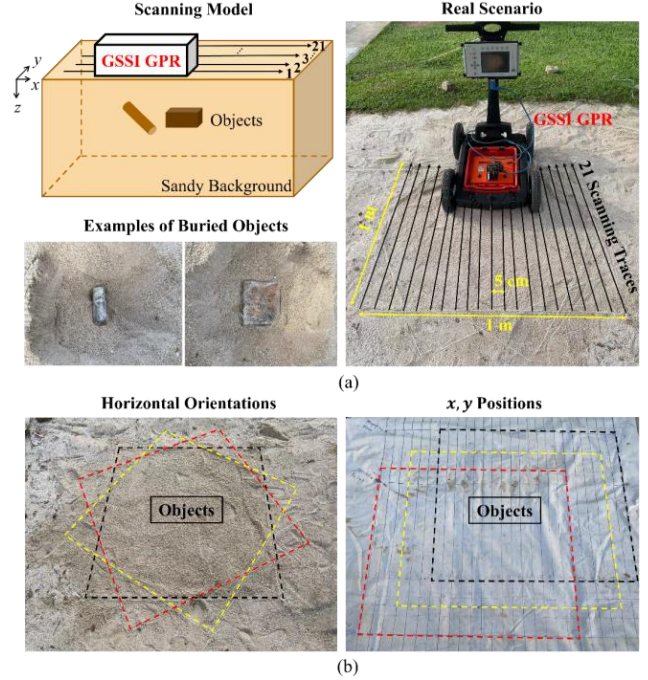

Fig. 9. The real experimental scenario. (a) The scanning model with a commercial GPR system. (b) The scanning areas for modelling various horizontal orientations (left) and $x, y$ positions (right).

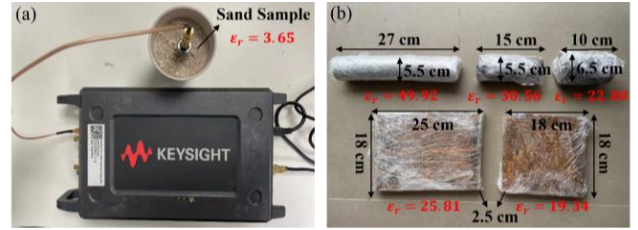

Fig. 10. (a) The relative permittivity measurement setup. (b) The properties of buried objects utilized in the real experiments.

Every measured C-scan is normalized to the range of [0, 1], resized to 128×128×128, and pre-processed by time-zero correction and mean subtraction [32] to form the input noisy C-scan. To obtain the ground-truth denoised C-scan, a C-scan is measured in an empty area far away from the object region and subtracted from the noisy C-scan. A total of 220 sets of real measured noisy C-scans, denoised C-scans, and the subsurface 3D permittivity maps are obtained to fine-tune the pre-trained Denoiser and Inverter for 100 epochs.

TABLE VII
EVALUATION METRICS FOR REAL MEASURED TESTING DATA

| | Groups | | i | | | | | | ii | | | | iii | | | i, ii, and iii |
|---|---|---|---|---|---|---|---|---|---|---|---|---|---|---|---|---|
| | Testing Data No. | (1) | (2) | (3) | (4) | (5) | Ave. | (6) | (7) | (8) | Ave. | (9) | (10) | Ave. | Ave. |
| Denoiser | SSIM (×10$^{-2}$) (↑) | 99.95 | 99.96 | 99.96 | 99.96 | 99.95 | **99.96** | 99.94 | 99.94 | 99.94 | **99.94** | 99.96 | 99.95 | **99.96** | **99.95** |
| | PSNR (dB) (↑) | 58.72 | 59.96 | 59.31 | 59.27 | 59.12 | **59.28** | 57.15 | 57.15 | 57.71 | **57.34** | 59.43 | 59.29 | **59.36** | **58.75** |
| | MAE (×10$^{-4}$) (↓) | 7.78 | 6.94 | 7.05 | 7.51 | 6.87 | **7.23** | 9.66 | 9.17 | 8.58 | **9.14** | 7.16 | 7.27 | **7.22** | **7.80** |
| | MRE (%) (↓) | 0.23 | 0.20 | 0.22 | 0.22 | 0.22 | **0.22** | 0.28 | 0.27 | 0.27 | **0.27** | 0.21 | 0.22 | **0.22** | **0.23** |
| Inverter | SSIM (×10$^{-2}$) (↑) | 99.64 | 99.60 | 99.76 | 99.70 | 99.61 | **99.66** | 99.57 | 99.57 | 99.48 | **99.54** | 99.44 | 99.37 | **99.41** | **99.57** |
| | MSE (×10$^{-1}$) (↓) | 4.89 | 8.52 | 1.73 | 2.50 | 1.25 | **3.78** | 6.58 | 3.54 | 8.57 | **6.23** | 8.81 | 11.37 | **10.09** | **5.78** |
| | MAE (×10$^{-2}$) (↓) | 3.38 | 4.31 | 2.26 | 2.73 | 2.69 | **3.08** | 4.15 | 3.85 | 5.14 | **4.38** | 5.59 | 5.79 | **5.69** | **3.99** |
| | MAPE (%) (↓) | 0.61 | 0.58 | 0.43 | 0.37 | 0.49 | **0.49** | 0.48 | 0.66 | 0.82 | **0.65** | 0.80 | 0.89 | **0.85** | **0.61** |

### B. Inversion Result Analysis

*1) Quantitative Results:* Using the fine-tuned networks, 10 measured testing data sets are used to evaluate the performance of the proposed scheme. The quantitative results of the Denoiser and Inverter are shown in Table VII. Groups i, ii, and iii include 5 one-object cases, 3 two-separated-object cases, and 2 two-touching-object cases, respectively. For the Denoiser, the average SSIM, PSNR, MAE, and MRE have an acceptable degradation compared to the previous numerical datasets. The accuracy degradation is reasonable as the real measured C-scans, as shown in Fig. 11(a), have randomly distributed noisy patterns interfering with the objects' reflection signatures. With the denoised C-scans, the reconstruction accuracy of the Inverter for various subsurface scenarios is evaluated in Table VII. It can be observed that the MSE, MAE, and MAPE of group i are the lowest and the SSIM of group i is the highest, while the four metrics of groups ii and iii are slightly worse than those of group i as the signatures of two objects in the obtained C-scan are more complicated. When two objects are in touch in group iii, their hyperbolic patterns are highly interleaved, resulting in a further decrease in reconstruction accuracy.

*2) Imaging Results:* The imaging results of denoised C-scans and reconstructed 3D permittivity maps are shown in Fig. 11. Cases 19-20 are one-object scenarios, Case 21 is a two-separated-object scenario, and Case 22 is a two-touching-object scenario. Fig. 11(a) presents the input noisy C-scans and the denoised C-scans obtained from the Denoiser. Selected B-scans that include the main reflection signatures or obvious noise patterns in each C-scan are visually presented. Compared to the noisy C-scans, the denoised ones contain less noise and clutters due to reflections from the subsurface environment, and the objects' reflection patterns are clearer. Fig. 11(b) shows the 3D reconstruction results predicted from the denoised C-scans using the Inverter. The predicted 3D permittivity maps and their ground truths, including the 3D geometries and one-slice of the permittivity maps, are compared. Cases 19 and 20 show the reconstruction results for a cylinder and a box, respectively. The results predicted by the Inverter match well with the ground truth ones. As for the two-separated-object scenario and the two-touching-object scenario shown in Cases 21 and 22, the reconstructed 3D permittivity maps are close to the ground truth ones in terms of the shapes, sizes, orientations, positions, and permittivity values of the subsurface objects. Although some clutters appear around the objects and the edge between the two touching objects is slightly blurry, the properties of the subsurface objects can be clearly observed. These quantitative and qualitative results demonstrate the applicability of the proposed 3DInvNet in effectively reconstructing the subsurface 3D permittivity maps from real measurement data.

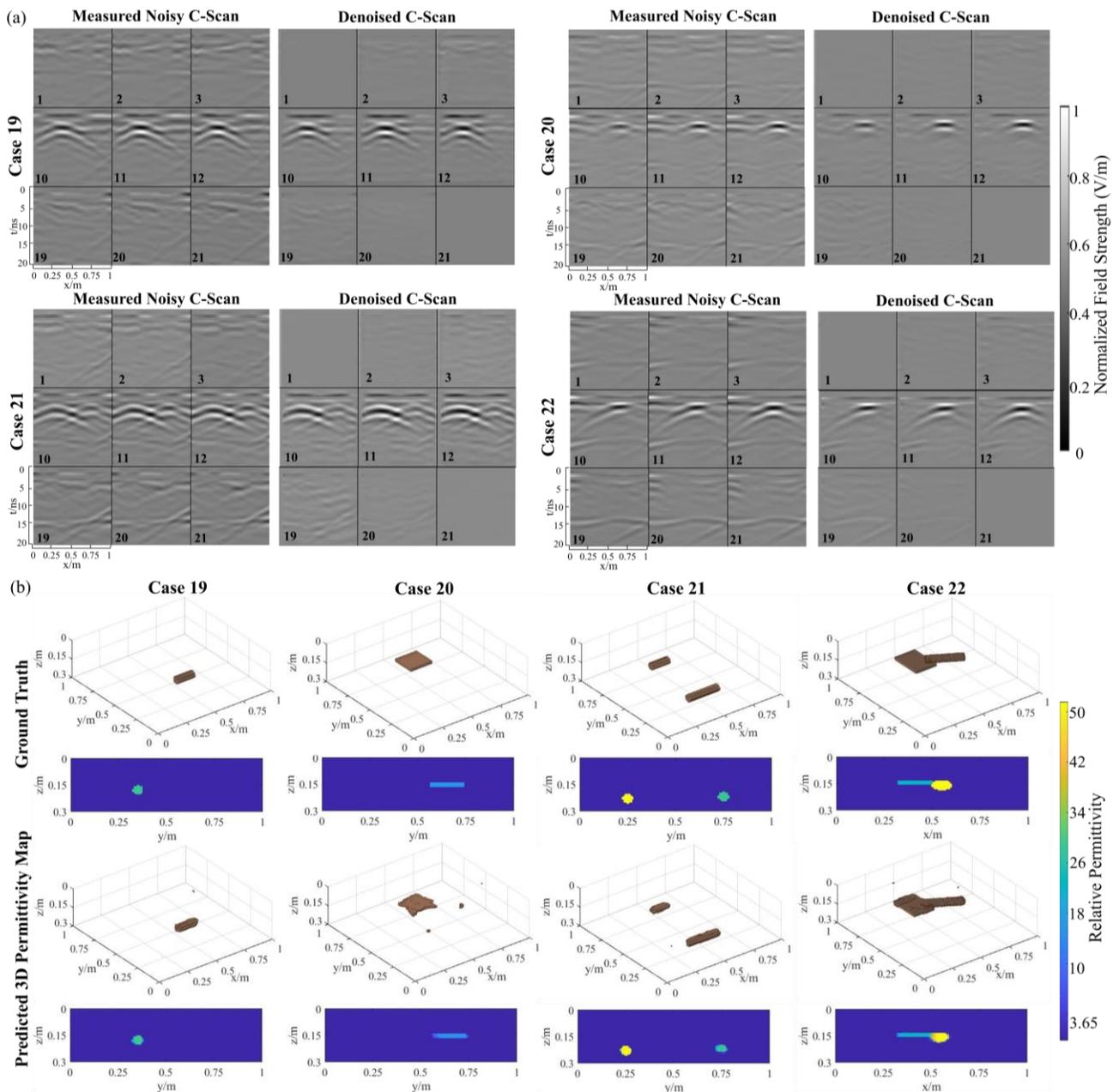

Fig. 11. (a) The 3D denoising results for real measured C-scans. (b) The 3D permittivity map reconstruction results. Cases 19-20 are one-object scenarios, Case 21 is a two-separated-object scenario, and Case 22 is a two-touching-object scenario.

## VI. Conclusion

In this paper, a two-stage scheme called 3DInvNet was proposed to rapidly denoise GPR C-scans and reconstruct 3D subsurface permittivity maps. The first stage, called Denoiser, effectively extracts the informative features of subsurface objects and suppress the environmental noise of the GPR C-scans. The second stage, called Inverter, builds the relationship between the discriminative features of the denoised C-scans with the corresponding subsurface scenarios. The test results on both the numerical and real measurement data demonstrate that the proposed method can accurately and efficiently reconstruct 3D permittivity maps. Moreover, the generalization capability and robustness of the proposed network were demonstrated by its application to various subsurface scenarios. This work was the first that implemented a 3D deep learning technique for the 3D subsurface structure restoration. It provides a novel and efficient method to solve the GPR 3D inverse problem and alleviate the computational burden of 3D reconstruction.

It should be noted here that the proposed scheme has certain limitations. 1) Supervised deep learning-based technique requires large datasets for pre-training the networks. The mapping relationships described by the networks trained with down-sampled datasets will be inaccurate. When the testing data are chosen slightly outside the pre-training set, transfer learning is used to fine-tune the pre-trained network with a small additional set of data. Nevertheless, when the testing cases are totally different from the scenarios in the pre-training set, the networks need to be retained with a large set of new data. 2) The current deep learning algorithms require trials to set the hyperparameters due to the immaturity of existing hyperparameter auto-tuning techniques. Our open-source framework is distributed at https://github.com/Qiqi-Dai/3DInvNet. The researchers can download and use our framework and improve it by implementing such auto-tuning algorithms in the future. 3) The tentative average permittivity of the soil medium is required for fine-tuning the inversion network. A diverse dataset considering different background environments can be helpful to address this issue, which will be investigated in our future work. Furthermore, developing physics-informed deep learning frameworks for GPR detection and imaging will be the focus of our future study.

References

[1] D. Feng, X. Wang, and B. Zhang, "Improving reconstruction of tunnel lining defects from ground-penetrating radar profiles by multi-scale inversion and bi-parametric full-waveform inversion," *Adv. Eng. Inform.*, vol. 41, Aug. 2019, Art. no. 100931.
[2] M. Pereira, D. Burns, D. Orfeo, Y. Zhang, L. Jiao, D. Huston and T. Xia, "3-D multistatic ground penetrating radar imaging for augmented reality visualization," *IEEE Trans. Geosci. Remote Sens.*, vol. 58, no. 8, pp. 5666–5675, Aug. 2020.
[3] L. Zou, Y. Wang, I. Giannakis, F. Tosti, A. M. Alani, and M. Sato, "Mapping and assessment of tree roots using ground penetrating radar with low-cost GPS," *Remote Sens.*, vol. 12, no. 8, Apr. 2020, Art. no. 1300.
[4] M. Wang, H. Wang, and H. Liu, "3D pre-stack reverse time migration of ground penetrating radar for subsurface imaging," in *Proc. 17th Int. Conf. Ground Penetrating Radar (GPR)*, Rapperswil, Switzerland, Jun. 2018, pp. 1–4.
[5] F. Lavoué, R. Brossier, L. Métivier, S. Garambois, and J. Virieux, "Two-dimensional permittivity and conductivity imaging by full waveform inversion of multioffset GPR data: A frequency-domain quasi-Newton approach," *Geophys. J. Int.*, vol. 197, no. 1, pp. 248–268, Apr. 2014.
[6] H. Qin, X. Xie, J. A. Vrugt, K. Zeng, and G. Hong, "Underground structure defect detection and reconstruction using crosshole GPR and Bayesian waveform inversion," *Autom. Constr.*, vol. 68, pp. 156–169, Aug. 2016.
[7] F. Watson, "Towards 3D full-wave inversion for GPR," in *Proc. IEEE RadarConf*, Philadelphia, PA, USA, May 2016, pp. 1–6.
[8] X. Wang and D. Feng, "Multiparameter full-waveform inversion of 3-D on-ground GPR with a modified total variation regularization scheme," *IEEE Geosci. Remote. Sens. Lett.*, vol. 18, no. 3, pp. 466–470, Mar. 2021.
[9] M. Salucci, M. Arrebola, T. Shan, and M. Li, "Artificial intelligence: New frontiers in real-time inverse scattering and electromagnetic imaging," *IEEE Trans. Antennas Propag.*, vol. 70, no. 8, pp. 6349–6364, Aug. 2022.
[10] X. Chen, Z. Wei, M. Li, and P. Rocca, "A review of deep learning approaches for inverse scattering problems (invited review)," *Prog. In Electromagn. Res.*, vol. 167, pp. 67–81, Jun. 2020.
[11] A. Massa, D. Marcantonio, X. Chen, M. Li, and M. Salucci, "DNNs as applied to electromagnetics, antennas, and propagation—A review," *IEEE Antennas Wirel. Propag. Lett.*, vol. 18, no. 11, pp. 2225–2229, Nov. 2019.
[12] Z. Wei and X. Chen, "Deep-learning schemes for full-wave nonlinear inverse scattering problems," *IEEE Trans. Geosci. Remote Sens.*, vol. 57, no. 4, pp. 1849–1860, Apr. 2019.
[13] O. Ronneberger, P. Fischer, and T. Brox, "U-Net: Convolutional networks for biomedical image segmentation," in *Proc. Int. Conf. Med. Image Comput. Comput.-Assisted Intervention*, Munich, Germany, Oct. 2015, pp. 234–241.
[14] L. Li, L. G. Wang, F. L. Teixeira, C. Liu, A. Nehorai, and T. J. Cui, "DeepNIS: Deep neural network for nonlinear electromagnetic inverse scattering," *IEEE Trans. Antennas Propag.*, vol. 67, no. 3, pp. 1819–1825, Mar. 2019.
[15] H. M. Yao, L. Jiang, and E. I. Wei, "Enhanced deep learning approach based on the deep convolutional encoder–decoder architecture for electromagnetic inverse scattering problems," *IEEE Antennas Wirel. Propag. Lett.*, vol. 19, no. 7, pp. 1211–1215, Jul. 2020.
[16] X. Ye, Y. Bai, R. Song, K. Xu, and J. An, "An inhomogeneous background imaging method based on generative adversarial network," *IEEE Trans. Microw. Theory Tech.*, vol. 68, no. 11, pp. 4684–4693, Nov. 2020.

[17] Z. Tong, J. Gao, and D. Yuan, "Advances of deep learning applications in ground-penetrating radar: A survey," *Constr. Build Mater.*, vol. 258, p. 120371, Oct. 2020.

[18] X. L. Travassos, S. L. Avila, and N. Ida, "Artificial neural networks and machine learning techniques applied to ground penetrating radar: A review," *Appl. Comput. Inform.*, vol. 17, no. 2, pp. 296–308, Apr. 2021.

[19] L. E. Besaw and P. J. Stimac, "Deep convolutional neural networks for classifying GPR B-scans," in *Proc. SPIE*, vol. 9454, May 2015, Art. no. 945413.

[20] W. Lei, J. Zhang, X. Yang, W. Li, S. Zhang, and Y. Jia, "Automatic hyperbola detection and fitting in GPR B-scan image," *Automat. Constr.*, vol. 106, Oct. 2019, Art. no. 102839.

[21] P. Bestagini, F. Lombardi, M. Lualdi, F. Picetti, and S. Tubaro, "Landmine detection using autoencoders on multipolarization GPR volumetric data," *IEEE Trans. Geosci. Remote Sens.*, vol. 59, no. 1, pp. 182–195, Jan. 2021.

[22] H. -H. Sun, Y. H. Lee, C. Li, G. Ow, M. L. M. Yusof, and A. C. Yucel, "The orientation estimation of elongated underground objects via multi-polarization aggregation and selection neural network," *IEEE Geosci. Remote. Sens. Lett.*, vol. 19, pp. 1–5, 2022.

[23] H. H. Sun, Y. H. Lee, Q. Dai, C. Li, G. Ow, M. L. M. Yusof, and A. C. Yucel, "Estimating parameters of the tree root in heterogeneous soil environments via mask-guided multi-polarimetric integration neural network," *IEEE Trans. Geosci. Remote Sens.*, vol. 60, pp. 1–16, 2022.

[24] J. K. Alvarez and S. Kodagoda, "Application of deep learning image-to-image transformation networks to GPR radargrams for sub-surface imaging in infrastructure monitoring," in *Proc. ICIEA*, Wuhan, China, May 2018, pp. 611–616.

[25] L. Xie, Q. Zhao, C. Ma, B. Liao, and J. Huo, "Ü-Net: Deep-learning schemes for ground penetrating radar data inversion," *J. Environ. Eng. Geophys.*, vol. 25, no. 2, pp. 287–292, Jun. 2021.

[26] B. Liu, Y. Ren, H. Liu, H. Xu, Z. Wang, A. G. Cohn, and P. Jiang, "GPRInvNet: Deep learning-based ground-penetrating radar data inversion for tunnel linings," *IEEE Trans. Geosci. Remote Sens.*, vol. 59, no. 10, pp. 8305–8325, Oct. 2021.

[27] Y. Ji, F. Zhang, J. Wang, Z. Wang, P. Jiang, H. Liu, and Q. Sui, "Deep neural network-based permittivity inversions for ground penetrating radar data," *IEEE Sensors J.*, vol. 21, no. 6, pp. 8172–8183, Mar. 2021.

[28] V. Badrinarayanan, A. Kendall, and R. Cipolla, "Segnet: A deep convolutional encoder-decoder architecture for image segmentation." *IEEE Trans. Pattern Anal. Mach. Intell.*, vol. 39, no. 12, pp. 2481–2495, Oct. 2016.

[29] P. Isola, J. Y. Zhu, T. Zhou, and A. A. Efros, "Image-to-image translation with conditional adversarial networks," in *Proc. IEEE Conf. Comput. Vis. Pattern Recognit. (CVPR)*, Honolulu, Hawaii, USA, Jul. 2017, pp. 5967–5976.

[30] H. Brunzell, "Detection of shallowly buried objects using impulse radar," *IEEE Trans. Geosci. Remote Sens.*, vol. 37, no. 2, pp. 875–886, Mar. 1999.

[31] A. M. Zoubir, I. J. Chant, C. L. Brown, B. Barkat, and C. Abeynayake, "Signal processing techniques for landmine detection using impulse ground penetrating radar," *IEEE Sens. J.*, vol. 2, no. 1, pp. 41–51, Feb. 2002.

[32] A. Benedetto, F. Tosti, L. B. Ciampoli, and F. D'Amico, "An overview of ground-penetrating radar signal processing techniques for road inspections," *Signal Process.*, vol. 132, pp. 201–209, Mar. 2016.

[33] F. Abujarad, G. Nadim, and A. Omar, "Clutter reduction and detection of landmine objects in ground penetrating radar data using singular value decomposition (SVD)," in *Proc. 3rd Int. Workshop Adv. Ground Penetrating Radar (IWAGPR)*, Delft, The Netherlands, May 2005, pp. 37–42.

[34] W. Xue, Y. Luo, Y. Yang, and Y. Huang, "Noise suppression for GPR data based on SVD of window-length-optimized hankel matrix," *Sensors*, vol. 19, no. 17, 2019, Art. no. 3807.

[35] B. Karlsen, J. Larsen, H. B. D. Sorensen, and K. B. Jakobsen, "Comparison of PCA and ICA based clutter reduction in GPR systems foranti-personal landmine detection," in *Proc. 11th IEEE Signal Process. Workshop Stat. Signal Process.*, Singapore, Aug. 2001, pp. 146–149.

[36] G. Chen, L. Fu, K. Chen, C. D. Boateng, and S. Ge, "Adaptive ground clutter reduction in ground-penetrating radar data based on principal component analysis," *IEEE Trans. Geosci. Remote Sens.*, vol. 57, no. 6, pp. 3271–3282, Jun. 2019.

[37] D. Kumlu and I. Erer, "Clutter removal in GPR images using nonnegative matrix factorization," *J. Electromagn. Waves Appl.*, vol. 32, no. 16, pp. 2055–2066, Jun. 2018.

[38] D. Kumlu and I. Erer, "Improved clutter removal in GPR by robust nonnegative matrix factorization," *IEEE Geosci. Remote Sens. Lett.*, vol. 17, no. 6, pp. 958–962, Jun. 2020.

[39] W. Lei, F. Hou, J. Xi, Q. Tan, M. Xu, X. Jiang, G. Liu, and Q. Gu, "Automatic hyperbola detection and fitting in GPR B-scan image," *Autom. Construct.*, vol. 106, Oct. 2019, Art. no. 102839.

[40] H. Liu, C. Lin, J. Cui, L. Fan, X. Xie, and B. F. Spencer, "Detection and localization of rebar in concrete by deep learning using ground penetrating radar," *Autom. Construct.*, vol. 118, Oct. 2020, Art. no. 103279.

[41] Q. Dai, Y. H. Lee, H. H. Sun, G. Ow, M. L. M. Yusof, and A. C. Yucel, "DMRF-UNet: A two-stage deep learning scheme for GPR data inversion under heterogeneous soil conditions," *IEEE Trans. Antennas Propag.*, vol. 70, no. 8, pp. 6313–6328, Aug. 2022.

[42] E. Temlioglu and I. Erer, "A novel convolutional autoencoder-based clutter removal method for buried threat detection in ground-penetrating radar," *IEEE Trans. Geosci. Remote Sens.*, vol. 60, pp.1–13, 2022.

[43] H. H. Sun, W. Cheng, and Z. Fan, "Learning to remove clutter in real-world GPR images using hybrid data," *IEEE Trans. Geosci. Remote Sens.*, vol. 60, pp. 1-14, 2022.

[44] Ö. Çiçek, A. Abdulkadir, S. S. Lienkamp, T. Brox, and O. Ronneberger, "3D U-Net: Learning dense volumetric segmentation from sparse annotation," in *Proc. Int. Conf. Med. Image Comput. Comput.-Assisted Intervention*, Athens, Greece, Oct. 2016, pp. 424–432.

[45] S. Anwar and N. Barnes, "Real image denoising with feature attention," in *Proc. IEEE Int. Conf. Comput. Vis.*, Seoul, Korea, Oct. 2019, pp. 3155–3164.

[46] K. He, X. Zhang, S. Ren, and J. Sun, "Deep residual learning for image recognition," in *Proc. IEEE Conf. Comput. Vis. Pattern Recognit. (CVPR)*, Las Vegas, Nevada, USA, Jun. 2016, pp. 770–778.

[47] J. Hu, L. Shen, and G. Sun, "Squeeze-and-excitation networks," in *Proc. IEEE Conf. Comput. Vis. Pattern Recognit. (CVPR)*, Salt Lake City, Utah, Jun. 2018, pp. 7132–7141.

[48] A. Araujo, W. Norris, and J. Sim, "Computing receptive fields of convolutional neural networks," *Distill*, vol. 4, no. 11, p. 151, Nov. 2019.

[49] D. P. Kingma and J. Ba, "Adam: A method for stochastic optimization," 2014, *arXiv:1412.6980*.

[50] C. Tan, F. Sun, T. Kong, W. Zhang, C. Yang, and C. Liu, "A survey on deep transfer learning," in *Proc. Int. Conf. Artif. Neural Netw.*, Rhodes, Greece, Oct. 2018, pp. 270–279.

[51] C. Warren, A. Giannopoulos, I. Giannakis, "gprMax: Open source software to simulate electromagnetic wave propagation for Ground Penetrating Radar," *Comput. Phys. Commun.*, vol. 209, pp. 163–170, Dec. 2016.

[52] C. Warren, C. Warren, A. Giannopoulos, A. Gray, I. Giannakis, A. Patterson, L. Wetter, A. Hamrah, "A CUDA-based GPU engine for gprMax: Open source FDTD electromagnetic simulation software," *Comput. Phys. Commun.*, vol. 237, pp. 208–218, Apr. 2019.

[53] N. R. Peplinski, F. T. Ulaby, and M. C. Dobson, "Dielectric properties of soils in the 0.3-1.3-GHz range," *IEEE Trans. Geosci. Remote Sens.*, vol. 33, no. 3, pp. 803–807, May 1995.

[54] I. Giannakis, A. Giannopoulos, and C. Warren, "A realistic FDTD numerical modeling framework of ground penetrating radar for landmine detection," *IEEE J. Sel. Top. Appl. Earth Obs. Remote Sens.*, vol. 9, no. 1, pp. 37–51, Jan. 2016.

[55] A. Paszke, S. Gross, F. Massa, A. Lerer, J. Bradbury, G. Chanan, T. Killeen, Z. Lin, N. Gimelshein, N. Gimelshein, A. Desmaison, A. Köpf, E. Yang, Z. DeVito, M. Raison, A. Tejani, S. Chilamkurthy, B. Steiner, L. Fang, J. Bai, S. Chintala, "Pytorch: An imperative style, high-performance deep learning library," in *Proc. Adv. Neural Inf. Process. Syst.*, Vancouver, Canada, vol. 32, Dec. 2019, pp. 8026–8037.

[56] I. Giannakis, A. Giannopoulos, and C. Warren, "Realistic FDTD GPR antenna models optimized using a novel linear/nonlinear full-waveform inversion," *IEEE Trans. Geosci. Remote Sens.*, vol. 57, no. 3, pp. 1768–1778, Mar. 2019.